\documentclass[aps,prl,reprint,longbibliography]{revtex4-2}

\usepackage{bm, amsmath, amsfonts, mathtools, physics}
\usepackage{ulem}
\usepackage{here}

\usepackage[dvipdfmx]{graphicx}
\usepackage[dvipdfmx]{color}
\usepackage{color}

\usepackage{graphicx}
\usepackage[font=small,labelfont=bf,justification=raggedright,singlelinecheck=false]{caption}

\usepackage{mathtools}
\usepackage{amssymb}
\begin{document}
    \title{Spatially Structured Flocking in a Proliferating Population of Self-Propelled Organisms without Explicit Alignment Interactions}

\author{Seiya Takamura${}^{1}$}

  \author{Nen Saito${}^{1}$ \email{nensaito@hiroshima-u.ac.jp}}

  \affiliation{  
  ${}^1$~Graduate School of Integrated Science for Life, Hiroshima University, 1-3-1 Kagamiyama, Higashi-Hiroshima City, Hiroshima, 739-8526, Japan 
  }

\begin{abstract}
While it is well established that self-propelled particles with alignment interactions can exhibit orientational order, the impact of self-replication and annihilation, which are key characteristics in cellular systems, on spatiotemporal order remains poorly understood. To explore the interplay between self-propulsion and self-replication, we introduce the active Brownian bug (ABB) model, in which self-propelled agents undergo stochastic, density-dependent replication and constant-rate death. Despite the absence of alignment interactions, the system exhibits flocking behavior characterized by high orientational order, while maintaining ordered hexagonal arrays. This emergent order arises from stochastic birth and death processes, offering a novel mechanism for flocking in  proliferating cellular populations.

\end{abstract}

    \maketitle
{\it Introduction:}
Motility and proliferation are fundamental features of living systems that support a wide range of biological functions. Collective migration observed in natural flocks and swarms has been theoretically studied using models of self-propelled agents~\cite{PhysRevLett.75.1226,baconnier2025self}, revealing that self-propulsion and alignment interactions are crucial for the emergence of polar orientational order. However, these models rarely incorporate agent replication or death, despite their importance in cellular systems.

On the other hand, spatiotemporal patterns in bacterial populations emerge through proliferation coupled with diffusion during colony formation. 
Bacterial colonies on agar plates have been observed to form radial arrays~\cite{budrene1991complex, budrene1995dynamics}, branching patterns~\cite{matsushita1998interface}, and more complex morphologies~\cite{ben1997snowflake}.
These collective behaviors have been modeled in the context of pattern formation using reaction-diffusion equations~\cite{matsushita1998interface, kawasaki1997modeling} and communicating random walkers~\cite{ben1994generic}.

The Brownian bug (BB) model~\cite{young2001reproductive} is a stochastic model that describes the aggregation patterns of bacteria and plankton. Despite the absence of interactions among agents, two simple rules, i.e., Brownian motion and a stochastic birth/death process, are sufficient to explain the emergence of aggregation of agents. Nearly identical models have also been proposed and analyzed in different contexts~\cite{iwasa1984branching, kuno1968studies}.
As an extension of the BB model, competitive interactions have been incorporated to explore the effects of local crowding or resource limitation~\cite{hernandez2004clustering, ramos2008crystallization, heinsalu2013clustering,martin2022coalescent,jorg2021stem,yamaguchi2017dynamical}. One such example is the neighborhood-dependent (ND) model~\cite{hernandez2004clustering}, which assumes that individuals compete for nutrients. In this model, the birth rate, which is constant in the BB model, depends on the number of neighboring agents. This density-dependent interaction leads to hexagonal crystal-like aggregation patterns~\cite{hernandez2004clustering, ramos2008crystallization, heinsalu2013clustering}.


While pattern formation driven by self-propulsion or proliferation has received considerable attention, the interplay between these two mechanisms has not been sufficiently explored. 
   This situation corresponds, for instance, to a proliferating population of self-propelled microorganisms capable of migrating or swimming~\cite{hallatschek2023proliferating,soares2019polarity}. However, in this study, we investigate pattern formation arising from the coexistence of proliferation and self-propulsion from a mathematical perspective, without assuming a specific biological system.

   To address this issue, we propose the active Brownian bug model, which incorporates self-propulsion into the ND model. Through extensive simulations, we identify a novel phase, referred to as the spatially structured flocking phase, in which aggregates arranged in a crystal-like hexagonal pattern collectively migrate in a unidirectional manner.
   Unlike typical models of active matter~\cite{PhysRevLett.75.1226,baconnier2025self}, our model includes no explicit alignment interactions in the dynamics of the self-propelling direction. 
Nevertheless, orientational order emerges through the stochastic birth/death process combined with competitive interactions, providing a fresh perspective on flocking scenario in growing populations of self-propelled agents.

  \begin{figure*}[thp]
		\includegraphics[keepaspectratio,width=\linewidth]{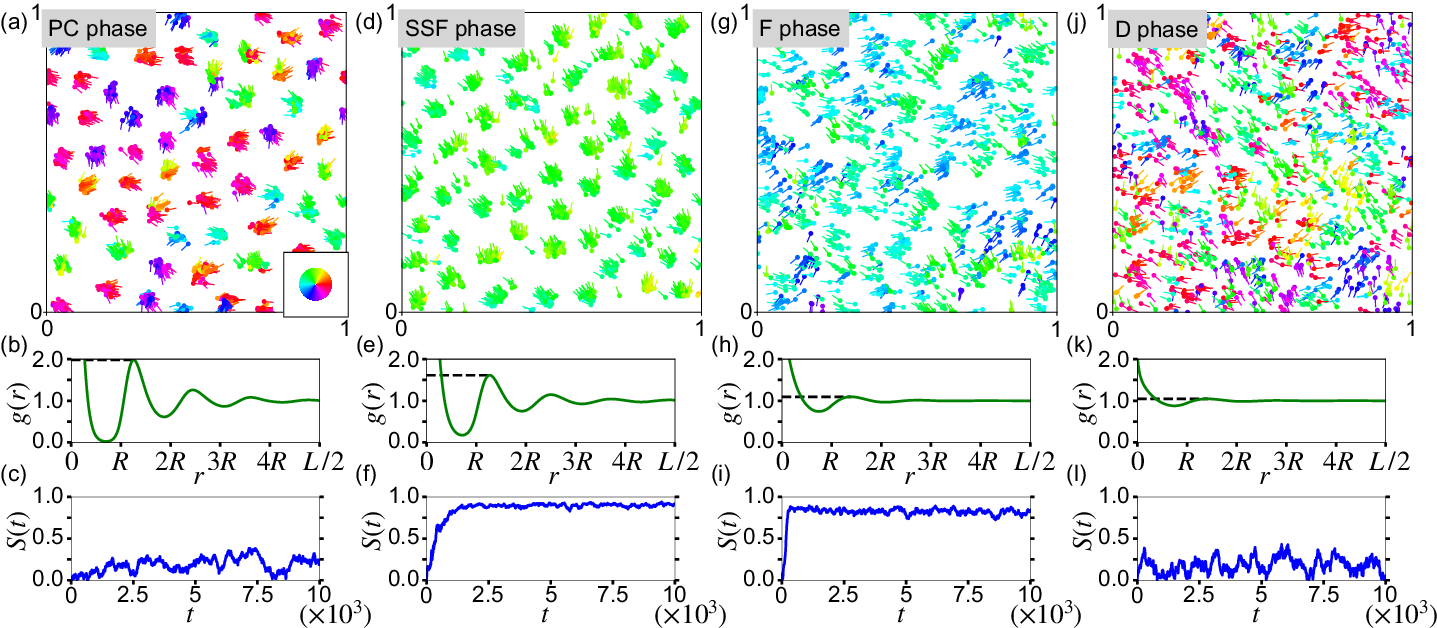}
		\caption{Spatiotemporal patterns in ABB model. 
        Panels (a), (d), (g) and (j) show simulation snapshots for PC ($D_r\simeq0.356\times10^{-5}$, $v_0\simeq0.356\times10^{-3}$), SSF ($D_r\simeq0.356\times10^{-5}$, $v_0\simeq3.557\times10^{-3}$), F ($D_r\simeq2\times10^{-5}$, $v_0\simeq11.25\times10^{-3}$) and D ($D_r\simeq6.325\times10^{-5}$, $v_0\simeq0.356\times10^{-3}$) phases, respectively. Panels (b), (e), (h), and (k) show the pair correlation functions $g(r)$, while the time series of the orientational order parameter $S$ are presented in panels (c), (f), (i) and (l). The other parameters are fixed as $D_\theta=2\times10^{-3}$, $L=1.0$, $p_0=0.85$, $q=0.15$, $R=0.1$, $N_s=50$, and $M_0=1500$.
        }
        \label{fig:snapshot}	
    \end{figure*}
    \par
    \vspace{1em}
{\it Model:}
    Here, we propose the active Brownian bug (ABB) model, in which self-propulsion is incorporated into the interacting Brownian bug model via a neighborhood-dependent reproduction rate~\cite{hernandez2004clustering}.    
    In this agent-based model, each agent $i$ is characterized by its position $\boldsymbol r_i=(x_i,y_i)$ and its self-propulsion direction $\theta_i$. The time evolution of these variables, reproduction, and death at each time step are determined by the following three procedures:
    \par
    \vspace{1em}
    \noindent
    (i) motion of agents
    \begin{eqnarray}
        \frac{dx_i}{dt}&=&\xi_{x_i}+v_0\cos\theta_i
        \\
        \frac{dy_i}{dt}&=&\xi_{y_i}+v_0\sin\theta_i
        \\
        \label{eq:dot_theta}
        \frac{d\theta_i}{dt}&=&\xi_{\theta_i}
    \end{eqnarray}
    where $v_0$ is the self-propulsion velocity. The noise terms $\xi_{x_i}$ and $\xi_{y_i}$ in the x-y coordinates represent white Gaussian noise with zero mean, $\langle\xi_{x_i}(t)\rangle=\langle\xi_{y_i}(t)\rangle=0$, and auto correlations $\langle\xi_{x_i}(t)\xi_{x_i}(t^\prime)\rangle=\langle\xi_{y_i}(t)\xi_{y_i}(t^\prime)\rangle=2D_r\delta(t-t^\prime)$. The noise $\xi_{\theta_i}$ in the self-propulsion direction is also white Gaussian noise with $\langle\xi_{\theta_i}(t)\rangle=0$ and $\langle\xi_{\theta_i}(t)\xi_{\theta_i}(t^\prime)\rangle=2D_\theta\delta(t-t^\prime)$. No explicit repulsion or attraction between agents is considered.
    \par
    \vspace{1em}
    \noindent
    (ii) update of the reproduction rate
    \par
The reproduction rate $p_i(t)$ of agent $i$ at time $t$ is determined by the following decreasing function of the number of neighboring agents $N_i$ located within a distance $R$ from $\boldsymbol r_i$:
    \begin{eqnarray}
        \label{eq:p_i}
        p_i(t)&=&\max\bigg\{p_0\bigg(1-\frac{N_i(t)}{N_s}\bigg),0\bigg\}
    \end{eqnarray}
    where $p_0$ is the maximum reproduction rate, and $N_s$ is a saturation parameter. This density dependence of the reproduction rate models the effect of local resource limitation, i.e., competition for nutrients among individuals, as assumed in the previous model~\cite{hernandez2004clustering}.  
 
    \par
    \vspace{1em}
    \noindent
    (iii) reproduction and death
    \par
    Reproduction and death occur based on the reproduction rate $p_i(t)$ and a constant death rate $q$, as follows:
    \begin{eqnarray}
        A&\xrightarrow{p_i}&A+A
        \\
        A&\xrightarrow{q}&\emptyset
    \end{eqnarray}
    Whether agent $i$ reproduces, dies, or remains unchanged is determined with probabilities $p_i(t)dt$, $q\,dt$, and $1-p_i(t)dt-q\,dt$, respectively. The coordinate $\boldsymbol r_{i^\prime}$ and the direction $\theta_{i^\prime}$ of the newborn agent $i^\prime$ are inherited from the parent agent $i$, i.e., $\boldsymbol r_{i^\prime}=\boldsymbol r_{i}$ and $\theta_{i^\prime}=\theta_{i}$ . This inheritance of the cell polarity has been observed in {\it Tetrahymena} and {\it Paramecium}~\cite{soares2019polarity}.
    \par
    \vspace{1em}
The iteration of steps (i)-(iii) with the time increment $dt$ is performed on an $L\times L$ square simulation box with periodic boundary conditions. As the initial condition, the simulation starts with $M_0$ individuals whose positions $\boldsymbol r_i$ and orientations $\theta_i$ are randomly assigned. To quantify the presence or absence of spatial structure, we compute the pair correlation function $g(r)$ as follows:
    \begin{eqnarray}
        g(r)&=&\bigg\langle\frac1{C_{pair}}\frac{(L/2)^2}{(r+\Delta r)^2-r^2}\,h(r)\bigg\rangle
    \end{eqnarray}
    where $C_{pair}$ is the number of sampled pairs, $h(r)$ is the histogram of pairwise distances using bin size $\Delta r$, and the bracket $\langle\cdot\rangle$ denotes the time average. 
    This function takes unity for a uniformly distributed population, while it exhibits a damped oscillatory pattern when the population forms a hexagonal crystal-like array.
    \par
    \vspace{1em}

{\it Results:}
    Simulations by the proposed model were performed with time increment $dt=0.01$ for $10^6$ simulation steps. 
    Typical simulation snapshots are shown in Figs.\ref{fig:snapshot} (a,d,g,j). For small $D_r$ and $v_0$, a periodic spatial aggregation pattern is observed (Fig.\ref{fig:snapshot}a), consistent with the periodic clustered (PC) phase reported in a previous study~\cite{hernandez2004clustering}. This periodic pattern is reflected in the oscillating curve in the pair correlation function $g(r)$ (Fig.\ref{fig:snapshot}b) and a high second peak value $g(r_2)$, which we adopt as a measure of spatial order. The orientational order parameter $S(t)=|\sum_j^{M(t)}\exp i\theta_j|/M(t)$ for population size $M(t)$ at time $t$ remains low throughout the simulation (Fig.\ref{fig:snapshot}c), indicating the absence of orientational order in the self-propulsion direction.  For small $D_r$ and large $v_0$, we identify a novel phase where clusters arranged in a hexagonal array migrate collectively in the same direction (Fig.\ref{fig:snapshot}d). We refer to this phase as the spatially structured flocking (SSF) phase. In the SSF phase, both a periodic spatial pattern with a high $g(r_2)$ value (Fig.\ref{fig:snapshot}e) and orientational order with a high $S$ value (Fig.\ref{fig:snapshot}f)  coexist. For large $D_r$ and $v_0$, the orientational alignment occurs in the absence of a clear hexagonal spatial pattern (Fig.\ref{fig:snapshot}g). This reflects on a high $S$ (Fig.\ref{fig:snapshot}i) with a low $g(r_2)$  (Fig.\ref{fig:snapshot}h). We refer to this phase as the flocking (F) phase.
    For a large $D_r$ and a small $v_0$, neither spatial pattern (Figs.~\ref{fig:snapshot}j, k) nor orientational order (Fig.~\ref{fig:snapshot}l) is observed. This phase is referred to as the disordered (D) phase.

\begin{figure}[tbhp]
		\centering
		\includegraphics[keepaspectratio,width=\columnwidth]{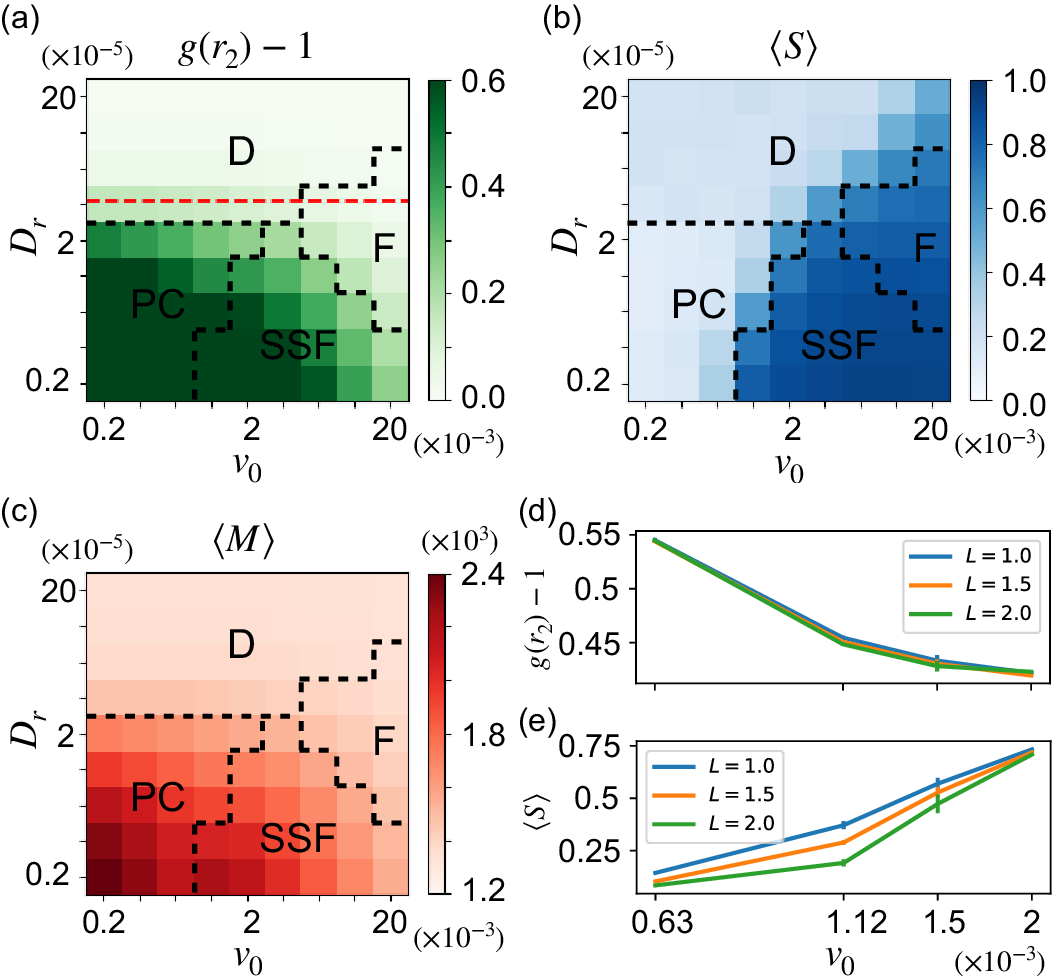}
		\caption{
            Phase diagrams in the $v_0-D_r$ parameter space. (a) Phase diagram of the second peak value, $g(r_2)$, of the pair correlation function $g(r)$. Darker colors indicate higher values. Panels (b) and (c) show the phase diagrams for $\langle S \rangle$ and $\langle M \rangle$, respectively. Time averages are computed from $5\times10^5$ to $10^6$ steps  with a sampling interval of 1000 steps. The examined $D_r$ and $v_0$ are $D_r=2\times 10^{-6+n/4} $ and $v_0=2\times 10^{-4+n/4} \ \ (n=0, 1, ..., 8)$. The parameters other than $v_0$ and $D_r$ are the same as fig.1.
            Panels (d) and (e) depict the dependency of $g(r_2)-1$ and $\langle S \rangle$ on $v_0$ at $D_r\simeq1.125\times 10^{-5}$ for different system sizes. The averages and standard errors were computed from 10 independent simulation runs. See also supplemental fig.~2.
        }
        \label{fig:phase}	
    \end{figure}

    These parameter dependencies are summarized in the phase diagram (Figs.\ref{fig:phase} a and b). The presence of the hexagonal pattern is judged by a heuristic criterion $g(r_2)>1.2$ (for PC and SSF phases), whereas the presence of the orientational order is determined by $\langle S\rangle>0.7$ (for SSF and F phases). The determination of phase boundaries using these criteria is somewhat ad hoc. Altering the threshold values leads to slight changes in the boundaries, however the parameter dependencies and the relative positions of each phase remain qualitatively unchanged (supplemental fig.~1). 
    As shown in Fig.\ref{fig:phase}a, the spatial pattern disappears for $D_r>2\times10^{-5} $, regardless of the $v_0$ value. This critical $D_r$ is consistent with the previous theoretical estimate in the ND model, $D_r\simeq(p_0-q)R^2/185.192=3.78\times10^{-5}$, indicated by the red dotted line in Fig.\ref{fig:phase}a. The mean population size at steady state, $\langle M \rangle$, in each parameter region is shown in Fig.\ref{fig:phase}c, which closely resembles $g(r_2)-1$ in Fig.\ref{fig:phase}a. This correspondence indicates that the presence of the hexagonal pattern enhances the carrying capacity.
    Throughout the entire examined parameter range, $\langle M \rangle$ in Fig.~\ref{fig:phase}c exceeds its mean-field approximation $\langle \tilde{M} \rangle=(1-q/p_0)N_sL^2/\pi R^2\simeq 1310$. The difference $\langle M \rangle -\langle \tilde{M} \rangle$ is shown in supplementary fig.2, indicating that the spatial structure consistently enhances the total population size, even in the disordered (D) phase.

    The orientational order parameter $\langle S \rangle$ increases sharply across the phase boundary from the PC phase to the SSF phase (Fig.\ref{fig:phase}b). We also confirmed that this increase in $\langle S \rangle$ becomes sharper as the system size $L$ increases (Fig.\ref{fig:phase}e), suggesting the flocking transition. On the other hand, $g(r_2)$ does not show significant changes for different $L$. Snapshots for $L=1.5$ and $2.0$ are shown in the supplemental fig.~2.

    \par
\begin{figure*}[thp]
		\centering
        \includegraphics[keepaspectratio,width=\linewidth]{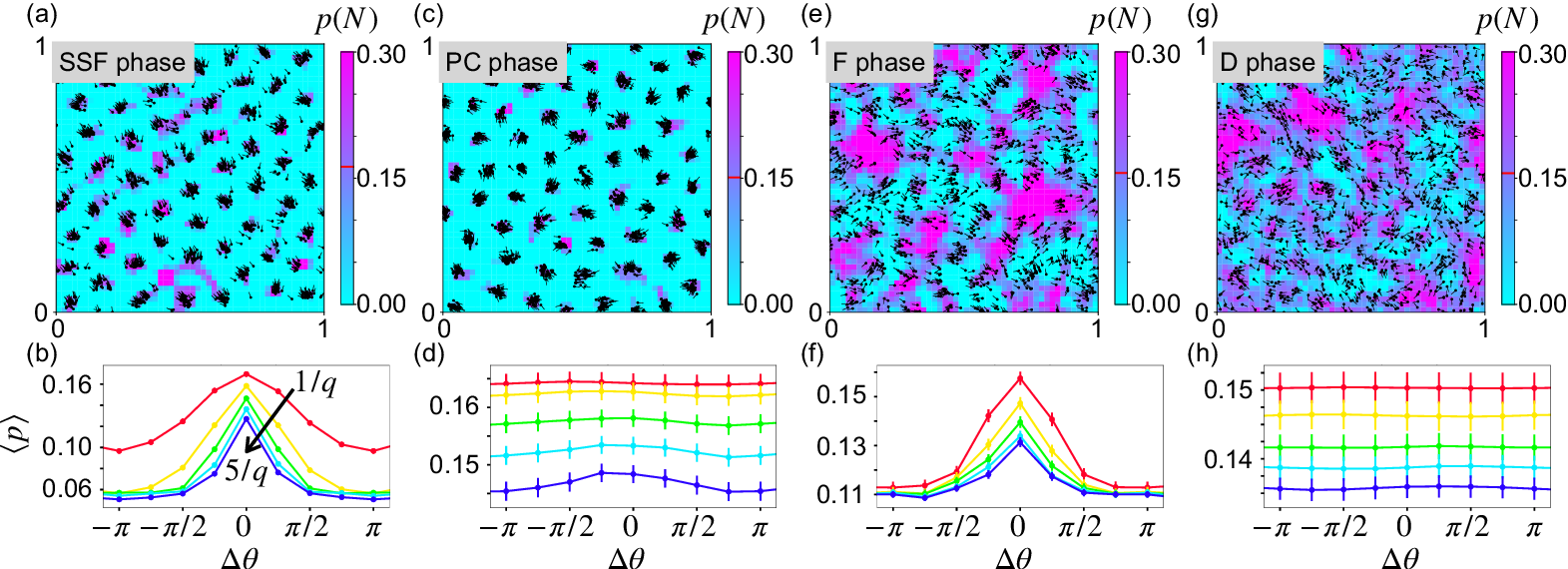}
		\caption{Visualizations of the proliferation rate and selection pressure. 
        Color maps in panels (a), (c), (e), and (g) show the spatial distribution of proliferation rates for each phase. Blue (red) regions indicate low (high) proliferation rates estimated using Eq.\ref{eq:p_i} based on the number of agents within a radius $R$ from the center of the box. Panels (b), (d), (f), (h) show the dependence of the proliferation rate on the self-propulsion direction. The average proliferation rate $\langle p\rangle$ over short time intervals $1/q$ to $5/q$ (top to bottom) is plotted against the deviation $\Delta\theta$ of the self-propulsion direction $\theta_i$ from the population average $\bar\theta$. Error bars represents standard errors. Details of the estimation are provided in the Supplemental Materials. Parameter settings for each phase are the same as those used for the corresponding phase in fig.~1.
        }
        \label{fig:dependency}	
    \end{figure*}

    The orientational order observed in the F and SSF phases is non-trivial, as the time evolution of $\theta_i$ in Eq.~\ref{eq:dot_theta} is driven purely by noise, with no alignment interaction introduced. To understand how this orientational order emerges, we quantified the spatial distribution of the reproduction rate in the SSF phase. Figure~\ref{fig:dependency}a shows spatial distribution of the reproduction rate, where the value at each spatial point is calculated based on the number of agents within a distance $R$ using Eq.~\ref{eq:p_i}. This figure indicates that reproduction rates are higher at the centers of clusters and lower in the empty regions between them. At first glance, this appears inconsistent with Eq.~\ref{eq:p_i}, which defines the reproduction rate as a decreasing function of local density. However, in the presence of a hexagonal configuration of aggregation, the empty regions contain several surrounding clusters within radius $R$, resulting in a higher local neighbor count than at the cluster centers, and thus a lower reproduction rate. Based on this spatial pattern, the maintenance of orientational order can be understood as follows: When the orientation of an agent at the center of a cluster deviates from that of its neighbors, the agent tends to leave the cluster and move into the empty region with a low reproduction rate. This process functions as a form of selection against misaligned agents, thereby maintaining the high orientational order.
This selective disadvantage of misalignment is demonstrated in Fig.~\ref{fig:dependency}b, which shows the direction-dependent reproduction rate averaged over a time window of $n/q$ ($n\times\text{average lifetime}$; $n=1,2,...,5$) for agents born at a given time (see Supplemental Materials). Here, $\Delta\theta$ in Fig.~\ref{fig:dependency}b represents the deviation of the self-propulsion direction from the population-averaged direction, defined as $\bar\theta=\arg(\sum_j^{M(t)}\exp i\theta_j)$. The result clearly demonstrates that deviation from $\bar\theta$ leads to a decrease in reproduction rate, which can be reduced by up to half for $\Delta\theta=\pm\pi$. Thus, the hexagonal spatial pattern imposes a negative selection pressure that contributes to the maintenance of orientational order in this phase.
    \par
    In the PC phase, a hexagonal configuration of clusters with elevated reproduction rates at their centers is also observed (Fig.~\ref{fig:dependency}c), while the selective disadvantage for misalignment disappears (Fig.~\ref{fig:dependency}d). Since $v_0$ is small in this phase, misaligned agents do not experience significant displacement from the cluster center within their lifetime. Despite the absence of selection pressure, the self-propulsion directions within each cluster tend to align. This can be attributed to the effect of neutral evolution in small populations in population genetics, which can homogenize the population without selection pressure.  
    Instead of the clear hexagonal pattern, the F phase exhibits scattered clusters with similar self-propulsion directions, surrounded by empty regions with a low reproduction rate (Fig.~\ref{fig:dependency}e). This situation resembles that of the SSF phase: misaligned agents are excluded as they drift into surrounding low-reproduction regions, which serve as a form of selection pressure maintaining orientational order (Fig.~\ref{fig:dependency}f). In the D phase, a similar spatial distribution of agents and reproduction rates to that in the F phase is observed (Fig.~\ref{fig:dependency}g), but orientational order, even within clusters, is disrupted. This disruption arises from strong perturbations associated with large $D_r$, which displace agents into empty areas or other clusters regardless of their alignment, thereby nullifying the reproductive advantage of the alignment  (Fig.~\ref{fig:dependency}h).    

{\it Discussion:}
In this study, we proposed the ABB model, which describes a growing population of self-propelled organisms, and discovered a novel form of collective unidirectional migration that maintains a ordered hexagonal spatial structure, which is specific to systems under the co-existence of self-propulsion and proliferation.
Notably, this flocking behavior emerges despite the absence of alignment interactions in the dynamics in the self-propulsion direction.
Our numerical analysis revealed that, as the self-organized aggregates in a hexagonal array pattern develops, regions of low growth rate spontaneously emerge between aggregates (Fig.~3a), effectively eliminating misaligned agents (Fig.~3b).
This spontaneously arising rectification mechanism is driven by selection pressure under density-dependent competitive interactions, and is clearly distinct from neutral effects in small populations known from population genetics, in which agents derived from a common ancestor occupy the entire population within a few generations, resulting in population homogenization.
Such neutral effects are likely related to the orientational alignment within clusters.

The emergence of the global orientational order in the present model relies on the assumption that a parent's cell polarity is inherited by its daughter cells. This assumption is exemplified by several biological systems. {\it Tetrahymena} and {\it Paramecium} are small unicellular organisms that swim by cilia. During cell division in the vegetative growth phase, the division plane is perpendicular to the anterior-posterior axis and the axis is inherited by the daughter cells~\cite{soares2019polarity}: the anterior structure of the parent is inherited by the anterior daughter cell, while the posterior structure is descended to the posterior daughter cell, indicating the inheritance of the polarity. Another example can be found in actin filament branching mediated by the Arp2/3 complex. Inside the cell, actin filaments branch when the Arp2/3 complex binds to them. If the actin treadmill is interpreted as a self-propelled agent, the branching process can be considered as proliferation of the agent. Although Arp2/3 induces branching with branching angle $\sim 70$ degree, cooperation with actin-bundling proteins that organize filaments into parallel bundles can result in daughter branches polymerizing in the same direction~\cite{ideses2008arp2}, which corresponds to the inheritance of the agent's self-propelled direction.

A previous study of the ND model revealed that the emergence of hexagonal arrays requires the condition $D<R^2(p_0-q)/185$ to be satisfied (red dotted line in Fig.~2a). This condition also holds in our model, even in the presence of self-propulsion. 
While the self-propulsion term can effectively increases the diffusion coefficient at sufficiently long timescales, the timescale of angular diffusion $1/D_\theta$ (= 500) is much larger than the agent's life time $1/q \simeq 6.7$. Therefore, the self-propulsion is essentially different from the diffusion within agent's lifetime and the agents cannot experience such enhanced diffusion. For this reason, the presence of the self-propulsion does not alter the condition of $D_r$ (the red dotted line in Fig.2). For a parameter condition $1/q > 1/D_\theta$, the effective diffusion needs to be taken into account and the red dotted line in Fig.2 would then behave as a decreasing function of $v_0$ in the $v_0$-$D_r$ plane. To maintain the hexagonal pattern, each cluster must be separated from neighboring clusters by at least a distance $R$; otherwise, agents interact competitively with members of multiple clusters, which significantly reduces the growth rate and disrupts the spatial structure. In addition, the formation of a stable hexagonal array requires that agents cannot travel between clusters during their lifetimes. This condition is satisfied in the PF and SSF phases, where the typical travel distance during an agent's lifetime $\tau \sim 1/q$ is approximately $v_0 \tau + \sqrt{2D_r \tau}$, which remains smaller than $R$.

The proposed ABB model focuses on a competitive population under local resource limitation, rather than on mechanically interacting agents~\cite{cates2015motility,almodovar2022liquid,bi2016motility,saito2024cell}, and does not account for repulsive interactions, which can alter the phase diagram as exemplified by the appearance of the motility-induced phase separation (MIPS)~\cite{cates2015motility}. Indeed, previous studies of proliferating self-propelled particles have reported the emergence of MIPS and the hexatic phases across a broad parameter range~\cite{almodovar2022liquid}, as well as extinction and coexistence in a binary mixture setting~\cite{almodovar2024extinction}. Such an extension with resource limitation needs to be addressed in future to understand the proliferating active matter systems~\cite{hallatschek2023proliferating,mishra2022active,cremer2019chemotaxis}.

In conclusion, our study demonstrates that a growing population of self-propelled organisms exhibits unidirectional collective migration while preserving a spatially structured pattern.
Although spatially structured collective migration has been reported in models with alignment interactions, such as propagating or localized bands~\cite{chate2008collective, kursten2020dry,barberis2016large} and flocking with crystalline order~\cite{das2024flocking,giavazzi2018flocking}, as well as in multi-species systems~\cite{dinelli2023non, oki2025anti, mandal2024robustness}, our results emphasize the role of the birth/death process in generating such behavior.
Clarification of the suggested flocking transition through a field-theory framework, as proposed in ~\cite{hernandez2004clustering, ramos2008crystallization}, could be an interesting direction for future work.

Acknowledgments: We acknowledge Masatoshi Ichikawa for helpful comments. This study was supported in part by the Japan Society for the Promotion of Science (JSPS) KAKENHI (21K03496, 23H04316, 25K07242 and 25H01364
to N.S.).

Data Availability Statement:
The codes and data used in this study are openly available on GitHub:~\cite{github2025}.

    \renewcommand{\refname}{References}
	\bibliographystyle{unsrt}

\end{document}